\documentclass[journal,transmag,twoside]{IEEEtran}
\ifCLASSINFOpdf
\else
\fi
\usepackage{amsmath,amsfonts}
\usepackage{graphicx}
\graphicspath{{./Figures/}}

\usepackage[capitalise]{cleveref}
\usepackage{booktabs}
\usepackage{tikz}
\usetikzlibrary{calc,arrows}
\usepackage{pgfplots}
\usepgfplotslibrary{colorbrewer}
\pgfplotsset{compat=newest}
\pgfplotsset{
    every axis/.append style={
    },
    /tikz/every picture/.append style={
        baseline
    }
}

\newcommand{\Matlab}{\textsc{Matlab}}
\newcommand{\CST}{CST}

\usepackage[math]{blindtext}
\usepackage{tablefootnote}

\begin{document}
%
%
\title{Gradient Coil Design and Realization\\ for a Halbach-Based MRI System}


\author{\IEEEauthorblockN{Bart de Vos\IEEEauthorrefmark{1},
Patrick Fuchs\IEEEauthorrefmark{1}~\IEEEmembership{Student Member,~IEEE},
Thomas O'Reilly\IEEEauthorrefmark{2}, \\
Andrew Webb\IEEEauthorrefmark{1,2}, and Rob Remis\IEEEauthorrefmark{1}}
\IEEEauthorblockA{\IEEEauthorrefmark{1}Circuits and Systems, Microelectronics, Delft University of Technology, Delft, The Netherlands}
\IEEEauthorblockA{\IEEEauthorrefmark{2}C.J.~Gorter Centre for High Field MRI, Radiology, Leiden University Medical Center, Leiden, The Netherlands}
\thanks{Manuscript received October 11th, 2019; revised December 5, 2019.
Corresponding author: P.S.~Fuchs (email: p.s.fuchs@tudelft.nl).}}

\markboth{IEEE Transactions on Magnetics,~Vol.~XX, No.~X, October~20XX}{De Vos~\MakeLowercase{\textit{et al.}}: Gradient Coils for a Halbach-Based MRI System}
%


\IEEEtitleabstractindextext{%
\begin{abstract}
In this paper we design and construct gradient coils for a Halbach permanent magnet array magnetic resonance (MR) scanner. The target field method, which is widely applied for the case of axial static magnetic fields, has been developed for a transverse static magnetic field as produced by a Halbach permanent magnet array. Using this method, current densities for three gradient directions are obtained and subsequently verified using a commercial magneto-static solver. Stream functions are used to turn the surface current densities into wire patterns for constructing the gradient coils. The measured fields are in good agreement with simulations and their prescribed target fields. Three dimensional images have been acquired using the constructed gradient coils with very low degree of geometric distortion.
\end{abstract}

}

\maketitle

\IEEEdisplaynontitleabstractindextext

%
\IEEEpeerreviewmaketitle

\section{Introduction}
%
%
%
\IEEEPARstart{G}{radient} coils are an integral part of MRI systems. Ideally, such coils produce linear magnetic fields that are used to spatially encode an object or body part: linearity allows simple image reconstruction via an inverse 2- or 3-dimensional Fourier transform ~\cite{Bernstein_etal}. Numerous methods for the design and optimisation of gradient coils have been proposed over the years~(e.g. \cite{Turner93,hidalgotobon10}), but most of these approaches are for conventional MRI scanners with the static magnetic field ($B_0$) aligned axially along the bore of the system.  

Interest in MRI in a low-resource setting is increasing \cite{Sarracanie_etal}. Conventional MRI hardware cannot be used under such circumstances, since it is expensive and generally difficult to maintain. Superconducting magnets, for example, are financially out of reach, and high power and fast switching requirements for gradient and radiofrequency hardware simply cannot be met. Moreover, conventional scanners are typically immobile and therefore cannot be easily transported to different locations.  

To address the difficulties that are encountered in a low-resource setting, new MR systems are being proposed such as MR scanners based on resistive magnets~\cite{Morrow,Zhang} or systems that utilize a Halbach permanent magnet array~\cite{Oreilly_etal, Cooley_etal}. For a resistive magnet, gradient coil design runs along similar lines as for conventional MRI systems, albeit typically for smaller bore sizes and lower power requirements. In contrast, for a Halbach array the background magnetic field is transverse to the bore as opposed to along the bore, and this provides additional challenges for the design of the gradient coils~\cite{Liu}. In a previous publication we described a 27 cm clear bore Halbach array designed ultimately for pediatric neuroimaging, operating at 2.15 MHz \cite{Oreilly_etal}. For this system simple non-optimized gradient coils were constructed, but the linear range was quite limited. 

In this paper, the target field method, as originally proposed by Turner \cite{Turner86}, is applied to design transverse oriented gradient fields. Specifically, a transverse gradient field is prescribed on an inner cylinder that is concentric to the Halbach array and the target field method is applied to find surface current densities on an outer cylinder that generate magnetic fields, which approximate this prescribed target field. Since this is an inverse source problem, regularization is required to obtain physically acceptable surface current densities. To this end, we follow the standard target field method and include regularization through apodization using a parametric spectral-domain Gaussian filter. By following this approach, $x$-, $y$-, and $z$- gradient coils are designed and realized. Furthermore, field simulations and measurements of these are presented, to show that the produced gradients are in good agreement with simulation, thereby verifying that the modified target field method can indeed be used to realize gradient coils in case the background field is transverse to the axis of the bore of a Halbach MR scanner. Finally, the gradient coils are incorporated in an experimental low-field Halbach MR scanner \cite{Oreilly_etal} thereby enabling us to use Fourier imaging techniques to acquire three-dimensional low-field MR images. Initial imaging results obtained with this scanner are presented as well. 

\section{Target Field Method}

\begin{figure}[tbp]
\centering
\includegraphics{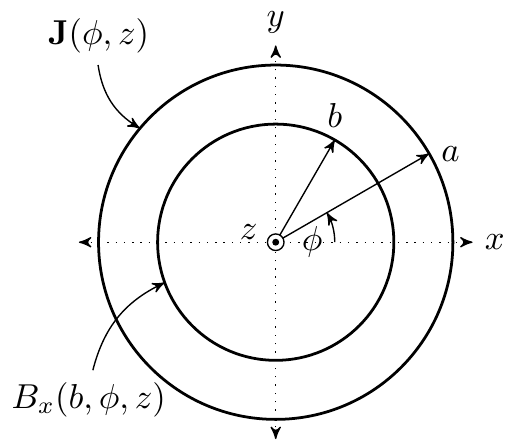}
\caption{Geometry and coordinate system for the target field method. The outer cylinder with radius $a$ supports a surface current $\mathbf{J}$ and the $x$-component of the magnetic field is prescribed on the inner cylinder with radius $b<a$. Both cylinders are of infinite length in the $z$-direction.}
\label{fig_geometry}
\end{figure}

To design gradient coils for a Halbach scanner with a transverse $B_0$ field, consider the cylindrical configuration illustrated in \cref{fig_geometry} consisting of two cylinders that extend to infinity in the positive and negative $z$-direction. The outer cylinder has a radius $a$ and the domains inside ($r<a$) and outside ($r>a$) the cylinder are filled with air. This cylinder supports a surface current density denoted by $\mathbf{J}$ and our objective is to find a surface current that approximates a prescribed magnetic field $\mathbf{B}$ on the inner cylinder with radius $b$. Given the cylindrical structure of our configuration, we will mainly work in cylindrical coordinates.

\begin{figure}[tbp]
  \centering
  \includegraphics{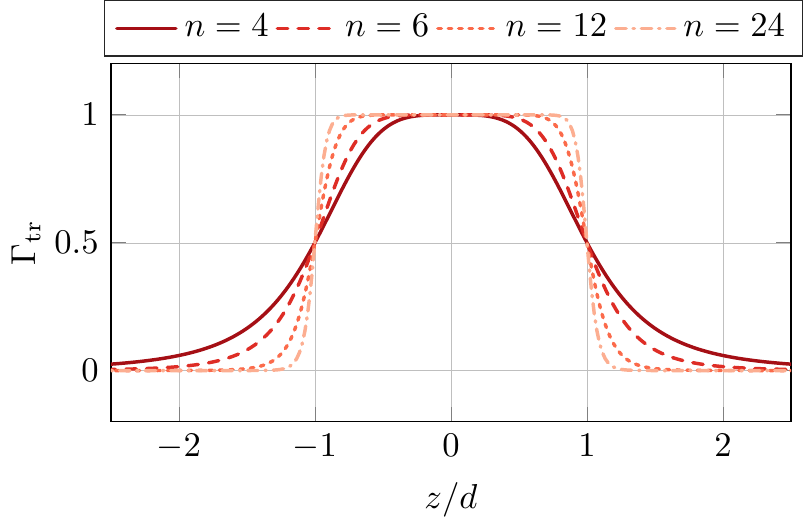}
  \includegraphics{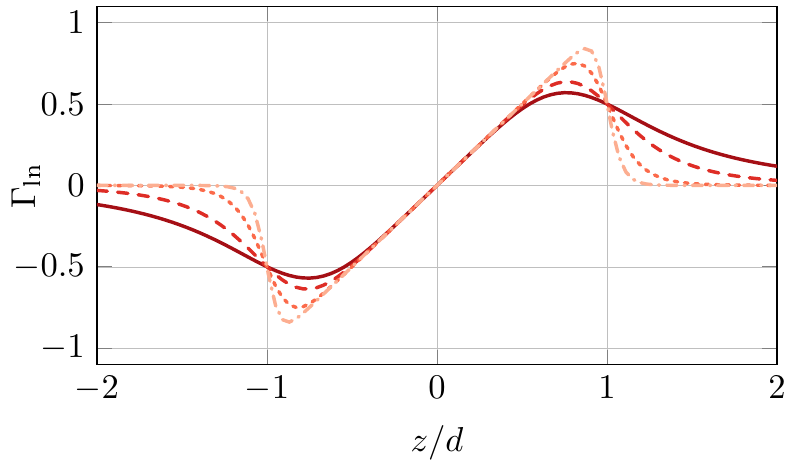}
  \caption{The gradient shape functions $\Gamma_\text{tr}$ (top) and $\Gamma_\text{ln}$ (bottom) as a function of $z/d$ for different values of the order $n$ of the gradient profile functions.}
  \label{fig_gradshapes}
\end{figure}

As a first step, we specify the target fields. Specifically, for a background field aligned in the $x$-direction, one of the following three linear $x$-directed gradient fields must be designed
\begin{equation}
\label{eq:Bx}
B_x(b,\phi,z) =
\begin{cases}
\Gamma_{\text{tr}}(z) b \cos(\phi) g_x \\
\Gamma_{\text{tr}}(z) b \sin(\phi) g_y \\
\Gamma_{\text{ln}}(z) g_z
\end{cases}
\end{equation}
on an inner cylinder with radius $b<a$ to derive surface currents (and ultimately the position of surface copper wires) that generate fields which approximate these target fields. In the above expressions, $g_{x,y,z}>0$ are constants and $\Gamma_{\text{tr}}(z)$ and $\Gamma_{\text{ln}}(z)$ are the transverse and longitudinal gradient shape functions given by
\begin{equation}
\label{eq:gamma_tr_gamma_ln}
\Gamma_{\text{tr}}(z) = \frac{1}{1+\left(\frac{z}{d}\right)^n} 
\quad \text{and} \quad
\Gamma_{\text{ln}}(z) = \frac{z}{1+\left(\frac{z}{d}\right)^n},
\end{equation}
respectively, where $d$ and $n$ ($n$ being an even integer) are tuning parameters that determine the length and decay rate of the gradient field in the $z$-direction. Note that $\Gamma_{\text{tr}}(z)$ is an even function of $z$, while $\Gamma_{\text{ln}}(z)$ is an odd function of $z$. Figure~\ref{fig_gradshapes} illustrates the two gradient shape functions as a function of $z/d$ for various choices of the order $n$.

To find a surface current density that approximately produces the prescribed target fields, we apply a two-dimensional Fourier transform with respect to the spatial coordinate $z$ and the angle $\phi$. For a generic field quantity $\Psi(r,\phi,z)$ this Fourier transform is given by
\begin{align}
\label{eq:FT_fw}
\tilde{\Psi}^{[m]}(r,k) &=
\int_{z=-\infty}^{\infty}
\int_{\phi=-\pi}^{\pi}
\Psi(r,\phi,z)
e^{-\text{j}m \phi} e^{-\text{j}k z}
\,
\text{d}\phi\, \text{d}z
\intertext{and the corresponding inverse Fourier transform is}
\label{eq:FT_inv}
\Psi(r,\phi,z) &=
\frac{1}{4\pi^2}
\int_{k=-\infty}^{\infty}
\sum_{m=-\infty}^{\infty}
\tilde{\Psi}^{[m]}(r,k)
e^{\text{j}m \phi} e^{\text{j}k z}
\, \text{d}k.
\end{align}
In Appendix~\ref{app_tfderiv} it is shown that the Fourier transform of the target field $B_x(b,\phi,z)$ is related to the Fourier transform of the $\phi$-component of the surface current by
\begin{align}
\label{eq:Bx_F}
\begin{split}
\tilde{B}_x^{[m]}(b,k) &=
\frac{\text{j}}{2}
\left[
\tilde{P}^{[m-1]}(b,k) - \tilde{Q}^{[m-1]}(b,k)
\right]
\tilde{J}^{[m-1]}_\phi(k)\\
&+
\frac{\text{j}}{2}
\left[
\tilde{P}^{[m+1]}(b,k) + \tilde{Q}^{[m+1]}(b,k)
\right]
\tilde{J}^{[m+1]}_\phi(k),
\end{split}
\end{align}
where $\tilde{P}^{[m]}(b,k)$ and $\tilde{Q}^{[m]}(b,k)$ are given by
\begin{align}
\label{eq:P}
\tilde{P}^{[m]}(b,k) &=  a\mu_0 k  I_m'(|k|b)K_m'(|k|a),
\intertext{and}
\label{eq:Q}
\tilde{Q}^{[m]}(b,k) &= m \frac{a\mu_0}{b}  \frac{|k|}{k}I_m(|k|b)K_m'(|k|a),
\end{align}
with $\mu_0$ the permeability of vacuum, and $I_m$ and $K_m$ modified Bessel functions of the first and second kind, respectively, and the prime indicates differentiation with respect to the argument of the Bessel functions. Note that $\tilde{P}^{[-m]}(b,k) = \tilde{P}^{[m]}(b,k)$ and $\tilde{Q}^{[-m]}(b,k) = - \tilde{Q}^{[m]}(b,k)$ for $m \in \mathbb{Z}$.

Since the target fields are known, equation (\ref{eq:Bx_F}) can be formally solved for the $\phi$-component of a spectral surface current density. However, similar to the standard target field method, such a current becomes unbounded as $|k| \rightarrow \infty$, which is not surprising, since we are attempting to directly solve an (ill-posed) inverse source problem. Therefore, regularization is applied in the form of a so-called apodization function $\tilde{T}(k)$, which serves as a low-pass filter that prevents exponential growth of the spectral domain current densities. Usually, the Gaussian function $\tilde{T}(k)=e^{-2(kh)^2}$ is used for apodization (with $h$ a regularization parameter) and we use this Gaussian in our approach as well.

Having found a solution to equation (\ref{eq:Bx_F}) and filtering out high spatial frequencies through multiplication by $\tilde{T}(k)$, the $\phi$-component of the surface current is obtained by substituting the filtered spectral solution into the inverse Fourier transform. Denoting the resulting spatial currents by $J_\phi^x$, $J_\phi^y$, and $J_\phi^z$ for the $\phi$-component of the surface current in the case of an $x$-, $y$-, or $z$-gradient target field, we obtain the surface current densities
\begin{align}
\label{eq:Jphi_x}
J_{\phi}^x(\phi,z)&=-\text{j} b \frac{g_x}{\pi} \cos(2\phi)\int_{k=-\infty}^{\infty}\frac{\tilde{\Gamma}_{\text{tr}}(k)\tilde{T}(k)}{\tilde{P}^{[2]}+\tilde{Q}^{[2]}}e^{\mathrm{j}kz}\mathrm{d}k,\\
\label{eq:Jphi_y}
J_{\phi}^y(\phi,z)&=-\text{j} b \frac{g_y}{\pi} \sin(2\phi)\int_{k=-\infty}^{\infty}\frac{\tilde{\Gamma}_{\text{tr}}(k)\tilde{T}(k)}{\tilde{P}^{[2]}+\tilde{Q}^{[2]}}e^{\mathrm{j}kz}\mathrm{d}k,
\intertext{and}
\label{eq:Jphi_z}
J_{\phi}^z(\phi,z)&=-\text{j}\frac{g_z}{\pi} \cos(\phi)\int_{k=-\infty}^{\infty}\frac{\tilde{\Gamma}_{\text{ln}}(k)\tilde{T}(k)}{\tilde{P}^{[1]}+\tilde{Q}^{[1]}}e^{\mathrm{j}kz}\mathrm{d}k.
\end{align}
The corresponding $z$-components of the surface current follow directly from the continuity equation. Further details can be found in Appendix~\ref{app_graddesign}.

Finally, from the computed current densities it is straightforward to extract the wire or current paths using stream functions as described in e.g.~\cite{Turner86}. These stream functions can then be used to realize the gradient coils. 

To verify our design method, we first compute the surface current densities given by (\ref{eq:Jphi_x}) -- (\ref{eq:Jphi_z}) and use stream functions to convert these current densities into wire patterns. These patterns are then used in a magnetostatic field solver to verify that currents flowing through the conductors of the gradient coils indeed produce the prescribed target fields. Subsequently, the three gradient coils were constructed and a magnetic field map of the $z$-gradient coil measured. Finally, the three gradient coils were incorporated into the low-field MRI Halbach-based scanner described previously. Three-dimensional imaging results obtained with this scanner are presented.  

\section{Results}

\subsection{Simulation results}
The surface current densities of (\ref{eq:Jphi_x}) -- (\ref{eq:Jphi_z}) were computed using \Matlab\footnote{Matlab 2018b, The MathWorks, Inc., Massachussets, USA.}. The regularization parameter was chosen as $h=0.05$ and the order $n$ of the target fields was taken as $n=16$ for the $z$-gradient coil and $n=30$ for the $x$- and $y$-gradient coils. These values were chosen in order for the physical length of the coils to correspond to the system requirements (the length of the magnet is 50 cm, and the gradients are constrained to an length of 37 cm inside the magnet). The design of the $y$-gradient coil is equivalent to the design of the $x$-gradient coil, since $J_\phi^y(\phi,z) = g_y\, g_x^{-1} J_\phi^x(\phi-\pi/4,z)$, that is, $J_\phi^y(\phi,z)$ is a scaled and rotated version of $J_\phi^x$. 

Subsequently, the computed surface current densities were turned into discrete current paths using stream functions~\cite{Turner86}. These current paths then served as input for a magnetostatic field simulation using \CST\footnote{Computer Simulation Technology, 2019, 3DS SIMULIA, Johnston, RI, USA.}. The simulations provided a magnetic field, which could then be compared with the prescribed target field (\ref{eq:Bx}).

\begin{figure}[tbp]
  \centering
  \includegraphics{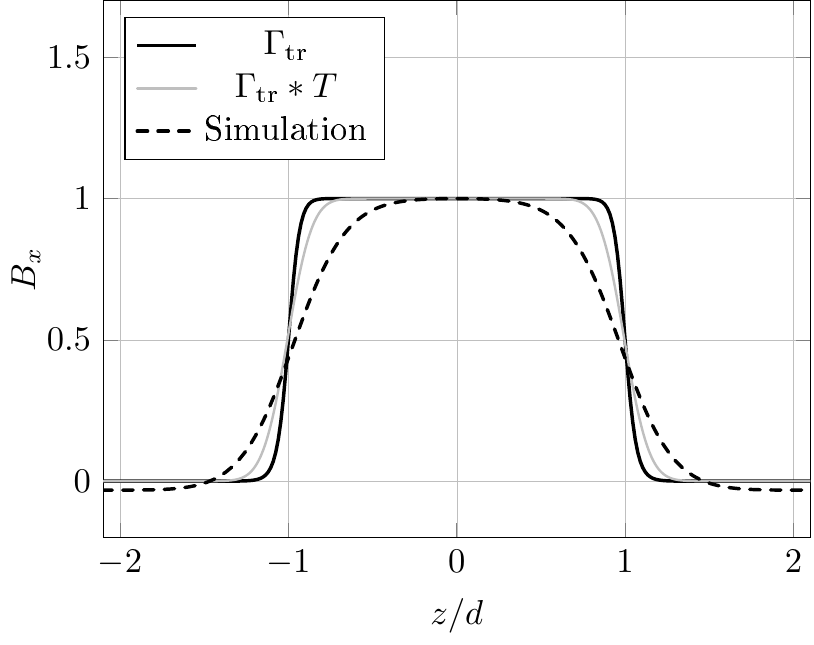}
  \caption{The $x$-gradient shape function along the bore of the coil (at radius $b$ and $\phi=0$) compared with the target field shape used to generate the gradient coil.}
  \label{fig_x_tf_vs_cst}
\end{figure}

\begin{figure}[tbp]
  \centering
  \includegraphics{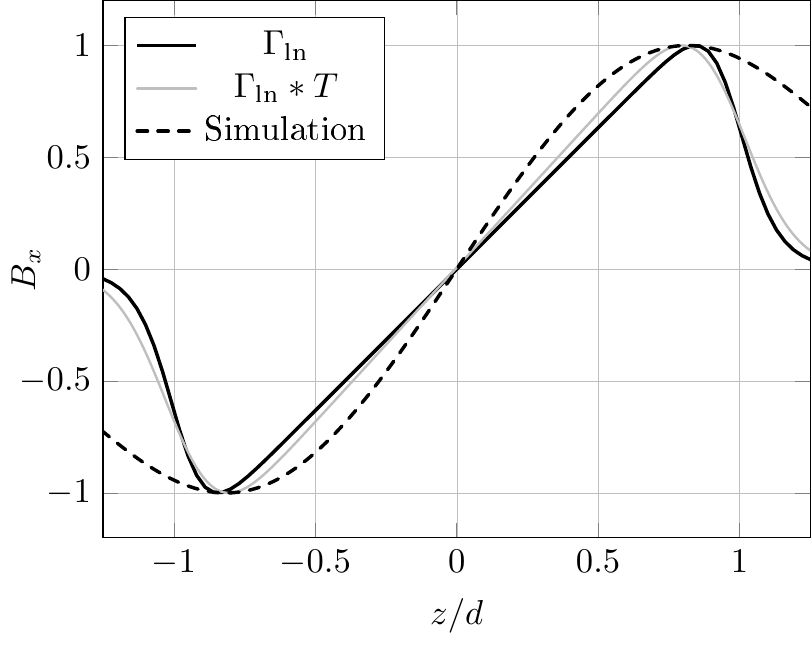}
  \caption{The $z$-gradient shape function along the bore of the coil (at radius $b$ and $\phi=0$) compared with the target field shape used to generate the gradient coil.}
  \label{fig_z_tf_vs_cst}
\end{figure}

This comparison can be found \cref{fig_x_tf_vs_cst,fig_z_tf_vs_cst}, where the prescribed target field profile functions are shown along with the simulated and normalized field along the bore of the coil at $\phi=0$ and $r=b$, since the target field is prescribed at this radius. As can be seen from the figures, the simulated fields closely follow the prescribed target profile functions. The difference is primarily caused by the apodization function. This function effectively smoothes the fields along the $z$-direction.


To study the effects of the coil parameters on the performance of the gradient coils, let us first consider the coil efficiency $\eta$, which is defined as the gradient strength produced by a unit current (T/m/A). We found that the order~$n$ of the target field profile function essentially does not influence the coil efficiencies of the $x$- and $y$-gradient coils. As $n$ increases, the distance between adjacent turns decreases, which will increase the inductance and shorten the physical coil length in the $z$-direction. The coil efficiency, however, remains essentially the same. On the other hand, $n$ does influence the efficiency of the $z$-gradient. For example, if we increase the order of the profile function from $n=6$ to $n=26$, the efficiency drops by approximately 20\%. Larger orders may be necessary for $z$-gradient coils, however, since otherwise the coil length in the $z$-direction may become longer than the length of the Halbach array. We note that special care must be taken when increasing $n$ in gradient coil design, since numerically the wires can be placed arbitrarily close together but in reality this is limited by the construction method. Close inspection of the current paths with the construction method in mind is needed to find the respective limits for this parameter.

\begin{figure}[tbp]
\centering
    \includegraphics[scale=0.8]{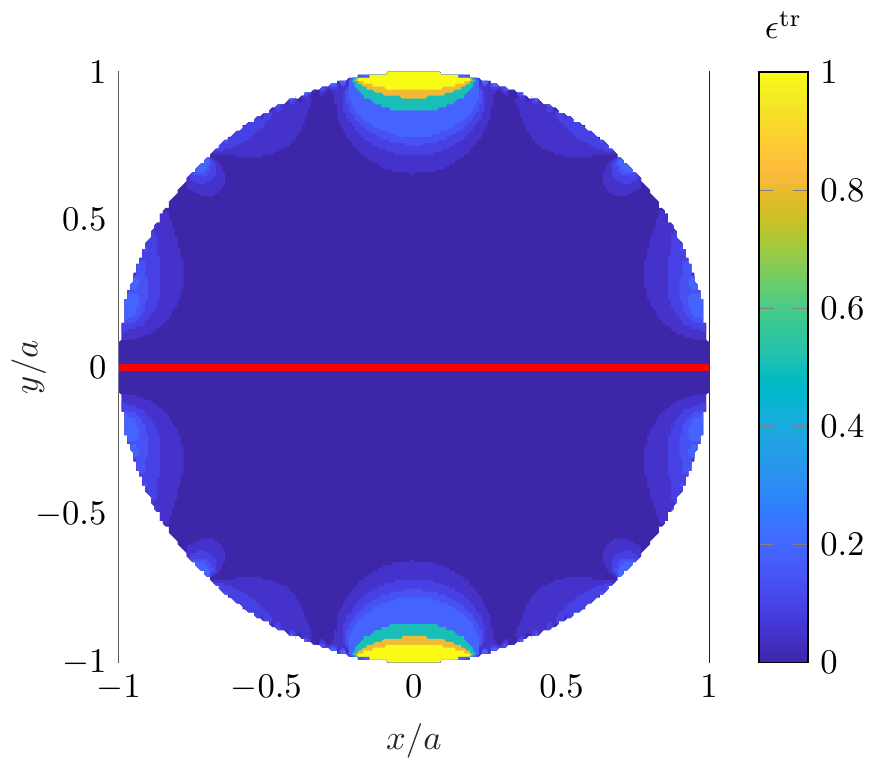}
    \includegraphics[scale=0.8]{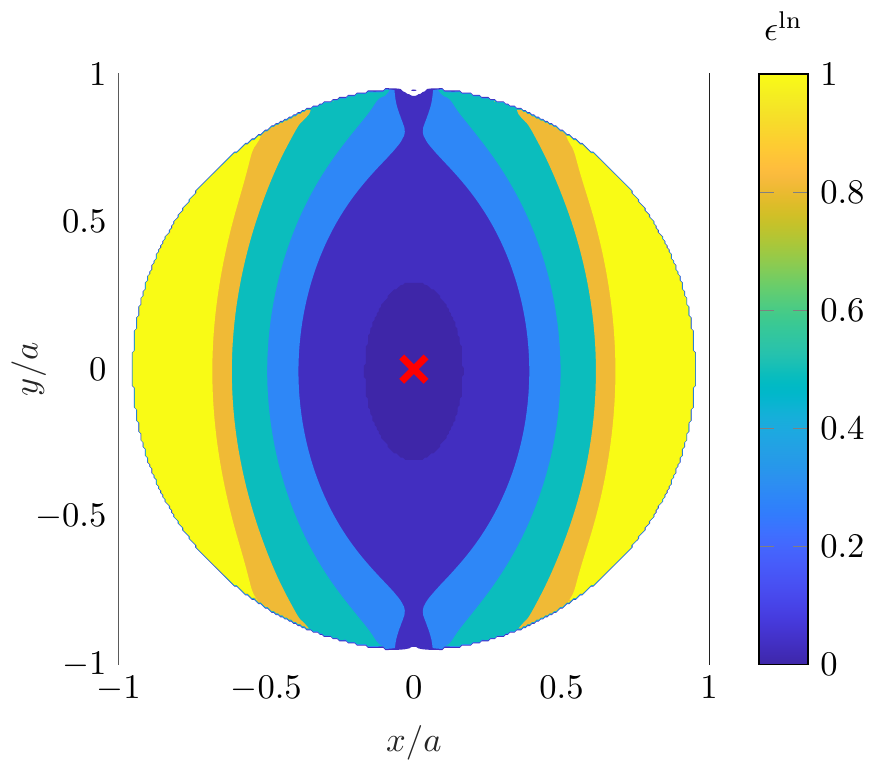}
\caption{Linear uniformity error $\epsilon^{\{\mathrm{tr},\mathrm{ln}\}}$ with respect to the field at the center of the bore.  The error $\epsilon^{\mathrm{tr}}$ for the $x$-gradient coil is shown at the top, while the error $\epsilon^{\mathrm{ln}}$ for the $z$-gradient coil is shown at the bottom. The red line and cross indicate the reference field line.}
\label{fig_err_trans}
\end{figure}

The linear uniformity of the gradient influences the region which can be imaged without distortions. This is quantified using the difference between the linear varying (prescribed) field and the field actually generated. In this case the simulated fields along the center line of the corresponding gradient are used as opposed to the prescribed fields. For the $x$-gradient this center line is across the bore in the $x$-direction (red line in Fig.~\ref{fig_err_trans}, top), for the $z$-gradient it is along the axis of the bore (cross in Fig.~\ref{fig_err_trans}, bottom). For the simulated field values $B_x$ this error is computed as 
\begin{equation}
\label{eq:Bx_error}
\epsilon^{\{\mathrm{tr},\mathrm{ln}\}}(x,y)=\frac{|B_x(x,y)-B_x^{\{\mathrm{tr},\mathrm{ln}\}}|}{|B_x^{\{\mathrm{tr},\mathrm{ln}\}}|}
\end{equation}
where the references $B_x^{\{\mathrm{tr},\mathrm{ln}\}}$ are defined as$B_x^{\mathrm{tr}} = B_x(x,0,0),$ and $B_x^{\mathrm{ln}}= B_x(0,0,z)$. These errors are displayed in \cref{fig_err_trans} from which it is immediately clear that the $x$-gradient field is  linear over a much larger area in the $xy$-plane than the $z$-gradient field. 

The uniformity of the gradient fields can also be described in terms of the linear spherical volume. Within this volume the deviation of the simulated field from a target field is less than 5\%. For the transverse $x$- and $y$ gradients the linear spherical volume is approximately 70\% of the diameter of the outer cylinder. In other words, a sphere centered at the origin and having a radius of 0.7$a$ completely encloses a region where the realized field deviates less than 5\% from the prescribed field. For the longitudinal $z$-gradient field, however, the linear volume is only 20\% of the diameter of the outer cylinder. Clearly, the linear region of the $z$-gradient coil is smaller than the linear region of the $x$- and $y$-gradient coils, which is due to the geometry of the $z$-gradient coil. In commercial scanners similar nonuniformity issues arise for these type of gradient fields and their effects in three-dimensional imaging are usually corrected in post-processing. 

To summarize, we have found that the coil efficiency~$\eta$ of the $x$- and $y$-gradient coils does not significantly vary for moderate changes in the order $n$ of the target field function. The coil efficiency of the $z$-gradient coil, however, is strongly dependent on $n$. Larger values of this parameter lead to $z$-gradient coils with a smaller spatial extent in the longitudinal $z$-direction, but decrease the coil efficiency. Moreover, for all coils the winding separation decreases as $n$ increases, which puts a restriction on the magnitude of the order~$n$ of the profile function, since in practice wires cannot be placed arbitrarily close to each other.     

\subsection{Gradient Construction}

To fixate and accurately position the wires, a 3D printable mould was created where the current paths were designed as slots. These slots facilitate easy and accurate placement of the wires. A single layer of 1.5 mm diameter copper wire was used to minimize the resistance and to reduce power dissipation. For the $z$-gradient coil, the order of the target field profile function $n=16$ is chosen, which leads to a gradient coil with a longitudinal length that is acceptable. For $x$- and $y$-gradient coils an order of $n=30$ was chosen. This was the maximum $n$ for which the adjacent wires (diameter $1.5$~mm) do not overlap. It must be noted that all three gradients have a slightly different radius as they are placed on top of each other. Denoting the radii of the $x$-, $y$-, and $z$-gradient coils by $a_x$, $a_y$, and $a_z$, respectively, the coils were placed on top of each other such that $a_z<a_y<a_x$. In other words, the $y$-gradient coil is positioned on top of the $z$-gradient coil, and the $x$-gradient coil on top of the $y$-gradient coil. This order of stacking was chosen because the $z$-gradient has the lowest efficiency and the $x$-gradient naturally has the highest performance due to the background field being  $x$-directed.

\begin{figure}[tbp]
\centering
\includegraphics[width=0.35\textwidth]{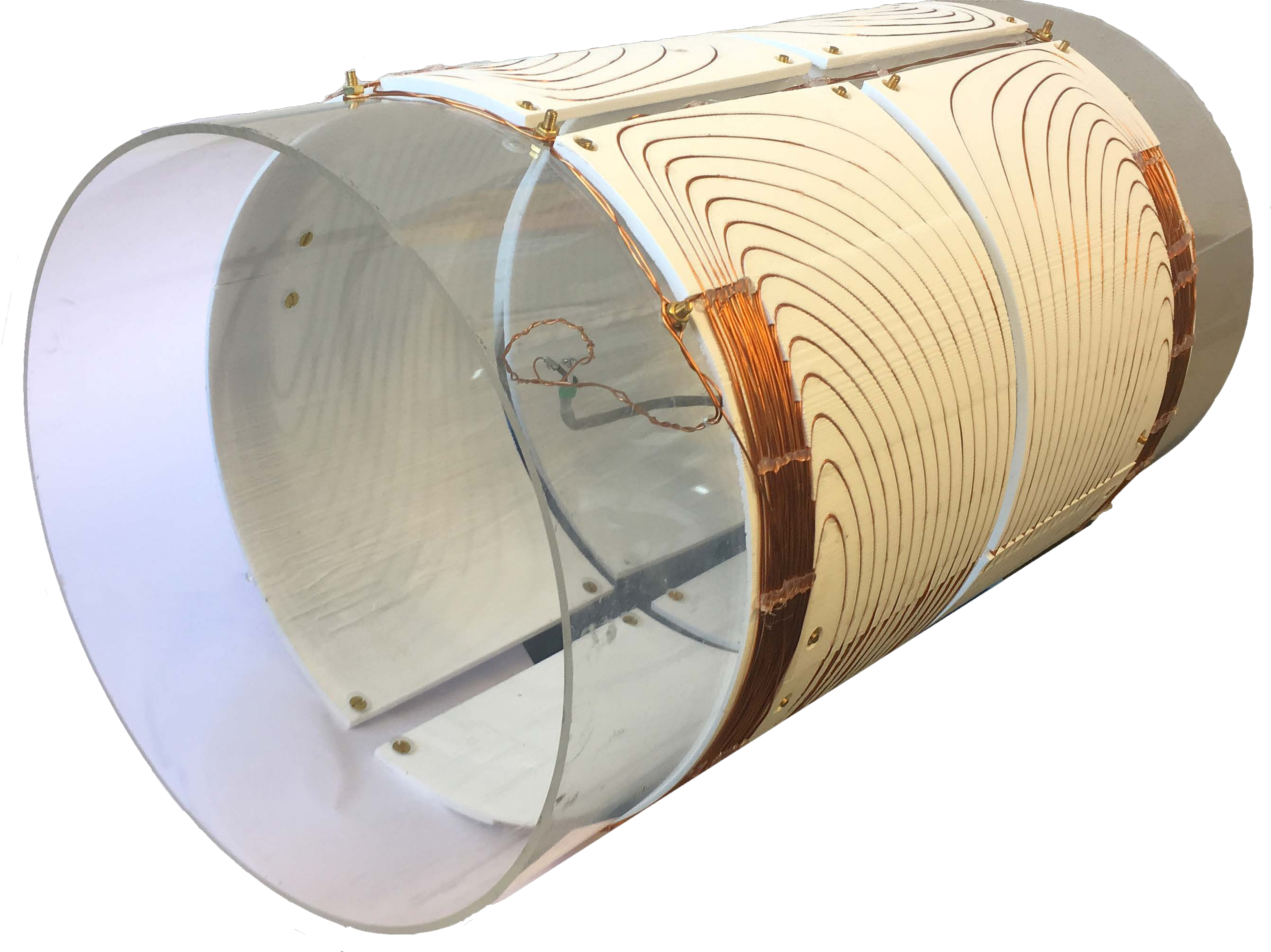}
\includegraphics[width=0.35\textwidth]{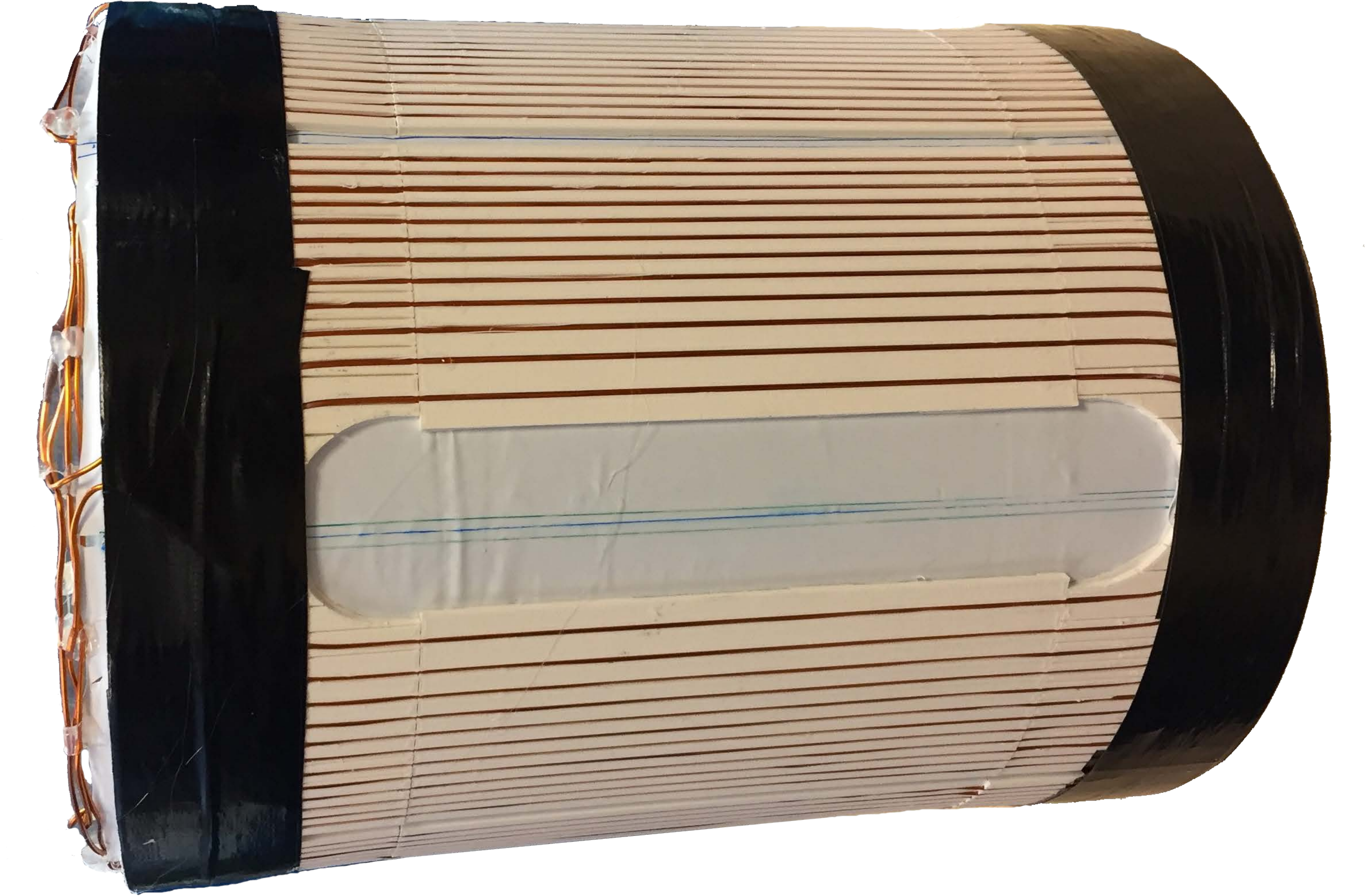}
\caption{A prototype z-gradient coil (top), where the 3D printed mould is clearly visible, and the gradient coil assembly after attaching the $y$-gradient (bottom).}
\label{fig_zgradient}
\end{figure}

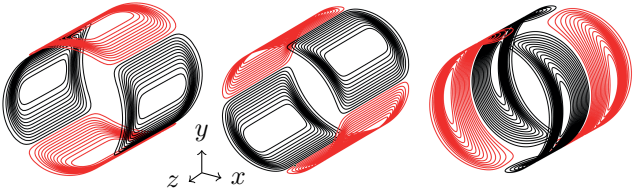
\begin{figure}[tbp]
\centering
\includegraphics[width=0.15\textwidth]{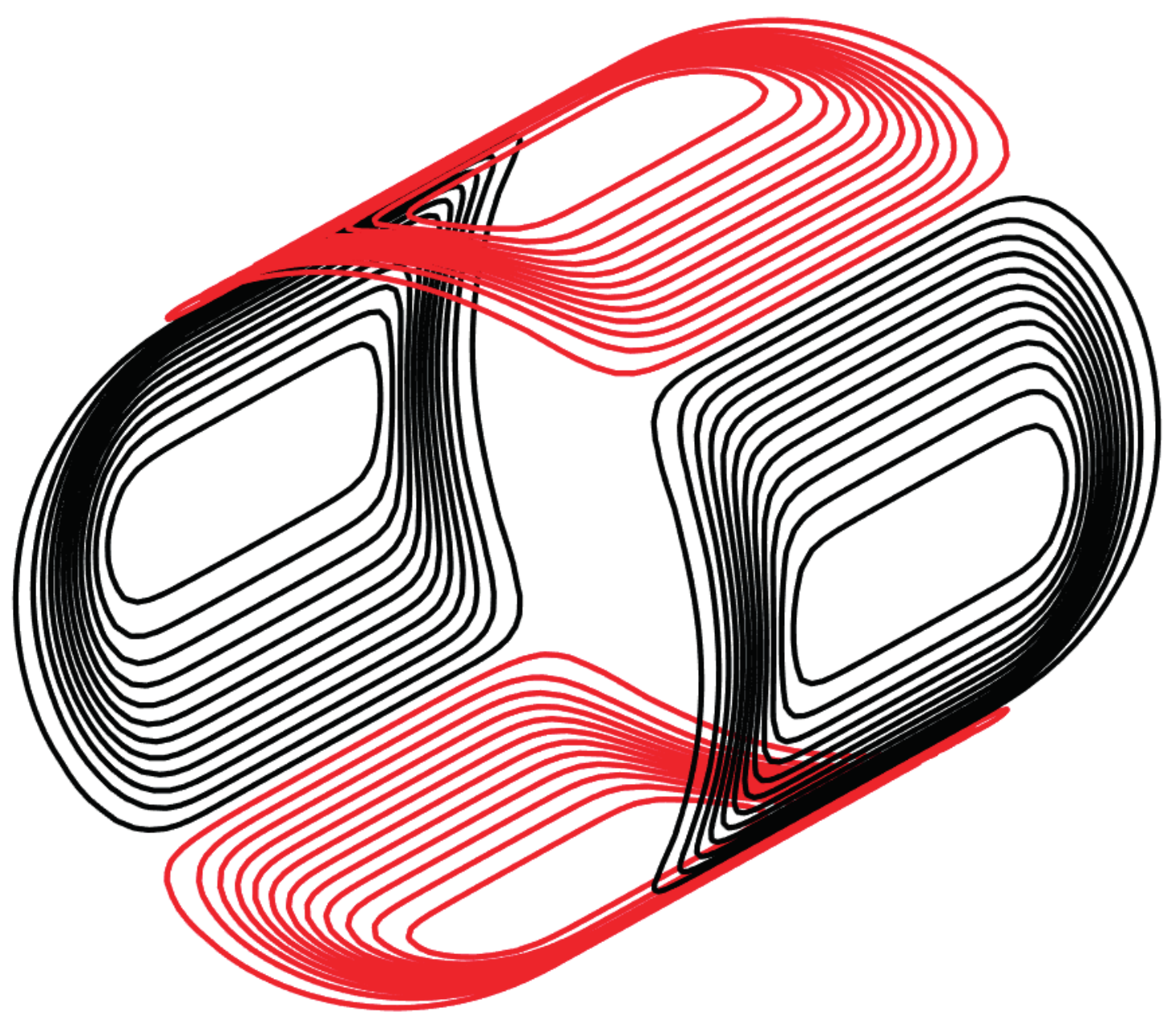}
\includegraphics[width=0.15\textwidth]{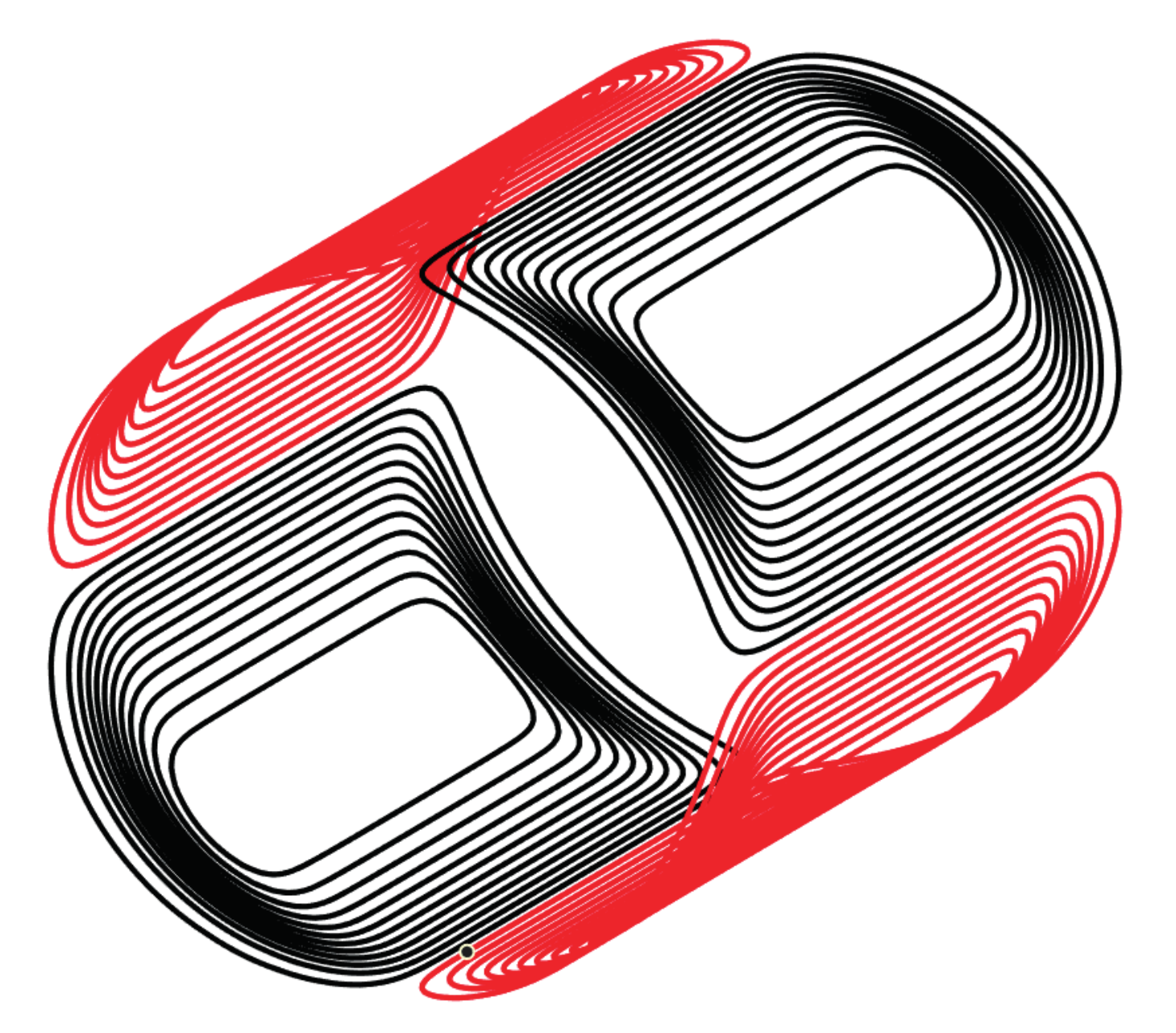}
\includegraphics[width=0.15\textwidth]{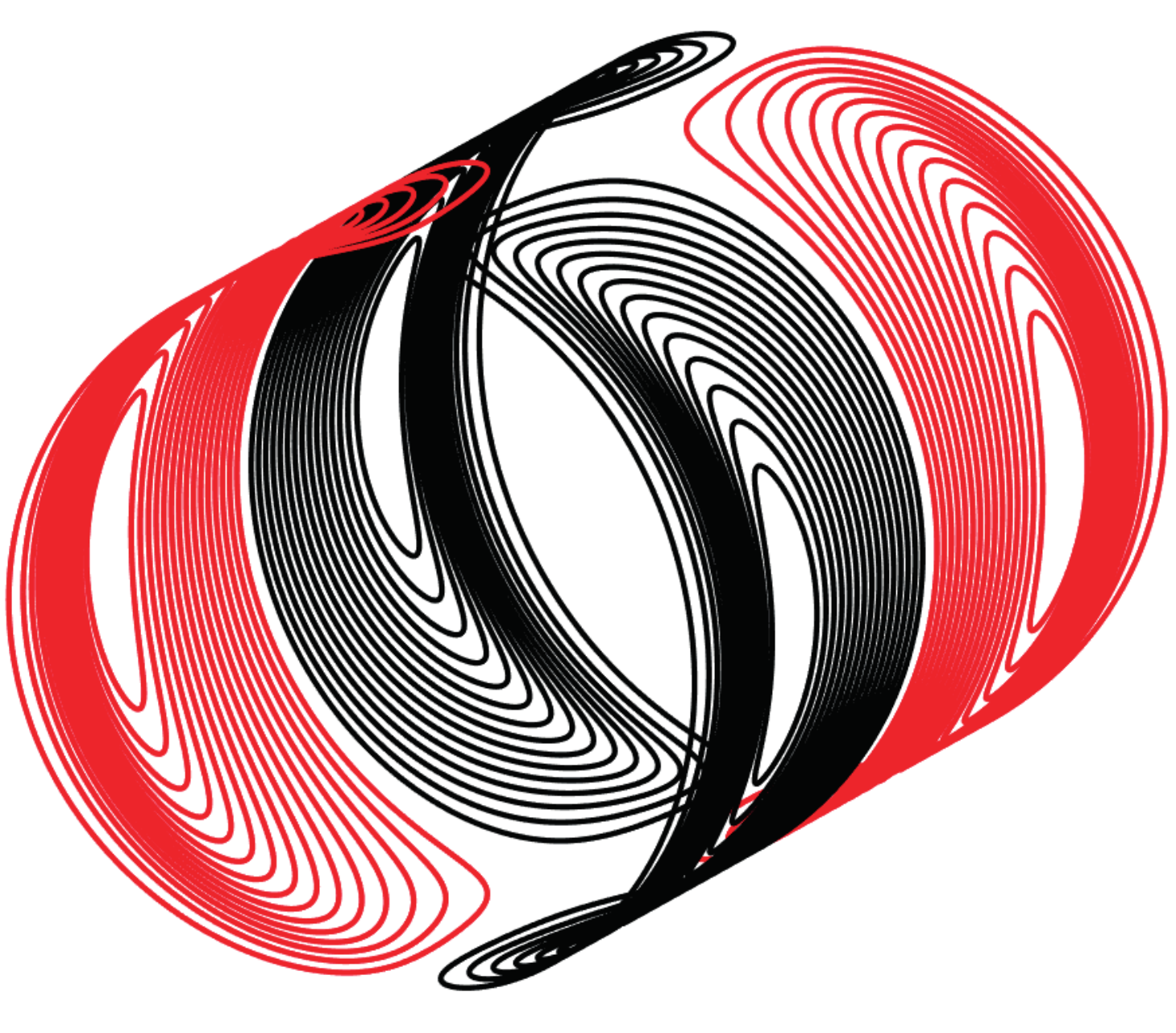}
 \begin{tikzpicture}[overlay,scale=0.3,xshift=-19.5cm,yshift=0.3cm] 
 \draw[->] (0,0) -- ({0.9*cos(-15)},{0.9*sin(-15)}) node[right] {$x$};
 \draw[->] (0,0) -- (0,1) node[above] {$y$};
 \draw[->] (0,0) -- ({0.7*cos(215)},{0.7*sin(215)}) node[left] {$z$};
 \end{tikzpicture}
\caption{Three-dimensional rendering of the wire paths of the $x$- (left), $y$- (middle), and $z$- (right) gradients. The color indicates the direction of the current: red for clockwise and black for counterclockwise currents.}
\label{fig_3d_gradients}
\end{figure}

The resistance and inductance of the coils were measured using a keysight U1733C RCL mete and a table of the coil design parameters and electrical properties can be found  in \cref{tab_tunparam}. Renderings of the wire paths of the coils are shown in \cref{fig_3d_gradients}, where currents run in a clockwise manner through the red wires and in a counterclockwise manner through the black wires. Lastly, photographs of the finished assembly and the 3D moulds can be found in \cref{fig_zgradient}.

\subsection{Measurements results}

\begin{figure}[tbp]
    \centering
    \includegraphics{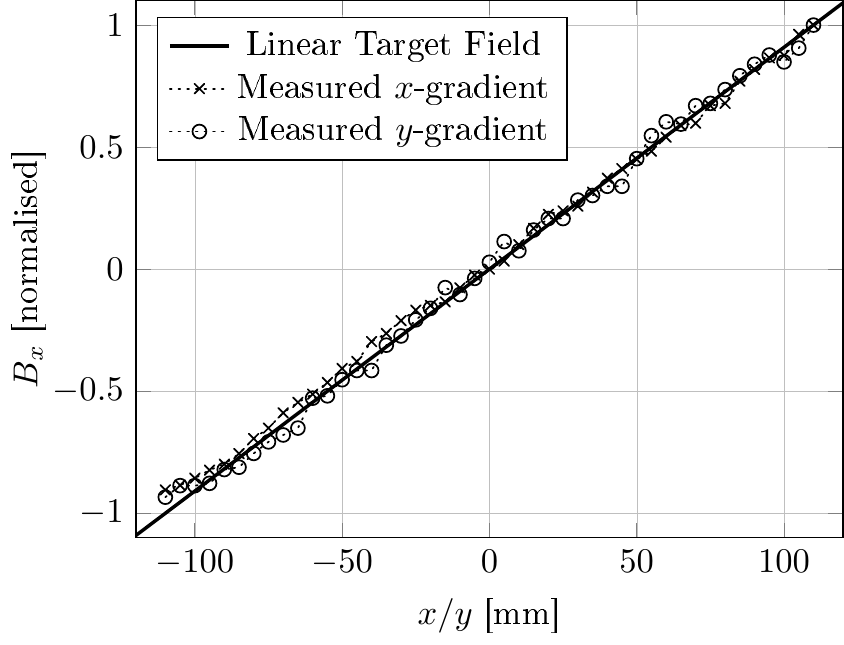}
    \caption{The simulated and measured $x$-component of the gradient field in a longitudinal slice through the center of a $x$- and $y$-gradient coil. The fields have been normalised for ease of comparison, the measured efficiency $\eta$ can be used to find the relation between current and field strength.}
    \label{fig_xymeas}
\end{figure}

\begin{figure}[tbp]
    \centering
    \includegraphics{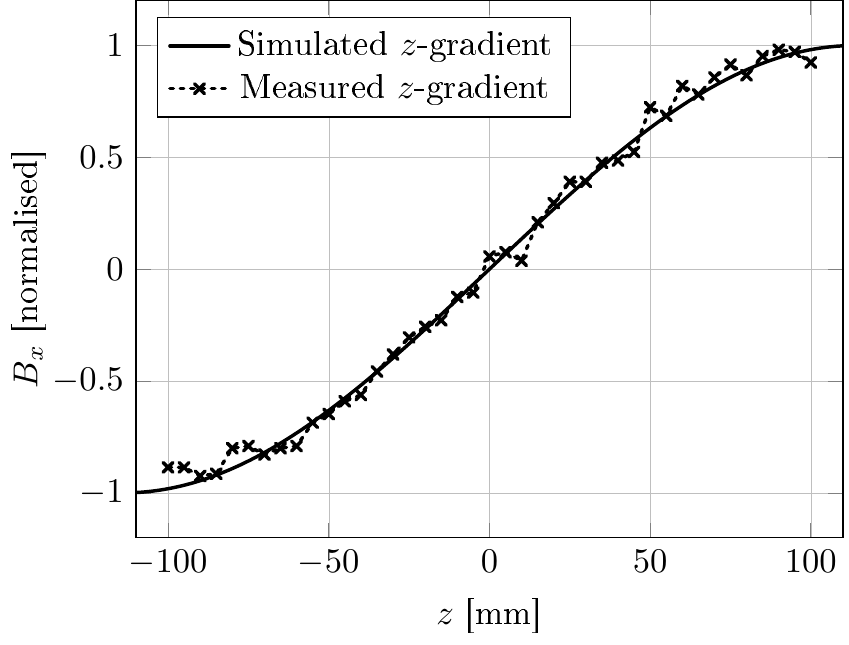}
    \caption{The simulated and measured $x$-component of the gradient field in a longitudinal slice through the center of a $z$-gradient coil. The fields have been normalised for ease of comparison, the measured efficiency $\eta$ can be used to find the relation between current and field strength.}
    \label{fig_zmeas}
\end{figure}

The field generated by the gradient coils is measured using a multipurpose 3-axis measuring robot \cite{measrobot}. The robot holds an AlphaLab inc. Gauss meter model GM2 which measures the field at a resolution of 10 mm isotropic. In \cref{fig_xymeas,fig_zmeas}, the $x$-component of the measured gradient field is shown as measured along the linear axis of the gradient at the center of the coil ($B_x(x,0,0)$ for the $x$-gradient, $B_x(0,y,0)$ for the $y$-gradient and $B_x(0,0,z)$ for the $z$-gradient). The measured and simulated fields are in good agreement with each other. Finally, for completeness we mention that the resistance and inductance of the coils were also measured using a keysight U1733C RCL meter. These can be found in \cref{tab_tunparam} together with the efficiency of the coils computed from the field measurements.

\begin{figure*}[htbp]
   \centering
    \includegraphics[width=0.32\textwidth]{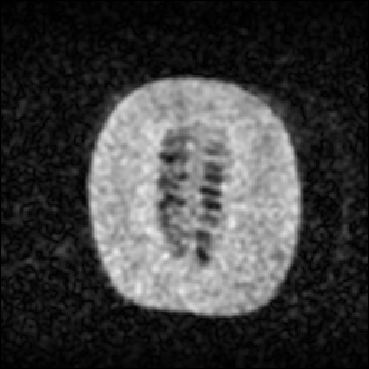}
    \includegraphics[width=0.32\textwidth]{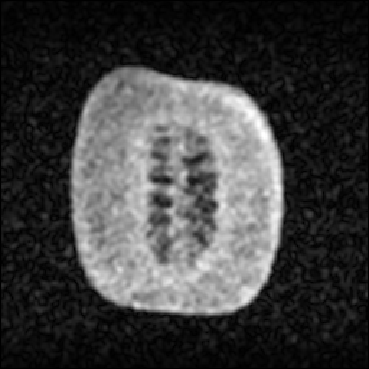}
    \includegraphics[width=0.32\textwidth]{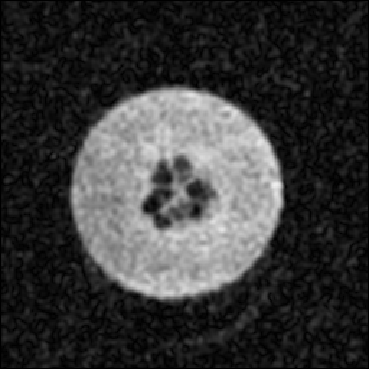}
   \caption{Coronal (left), sagital (middle), and transverse (right) images of a melon obtained with the low-field scanner of the LUMC that incorporates the gradient coils described in this paper. A 3D-TSE sequence was used with TE/TR = 30 ms/2000 ms. Echo train length was 32, with a Field of View of 192x192x192 mm. The data matrix consisted of 128 x 128 x 128 complex points. Total acquisition time was 17 minutes 4 seconds.}
   \label{fig_3d_scan_results}
\end{figure*}

The constructed coils were incorporated in an experimental low-field Halbach MR scanner that is currently under development at the Leiden University Medical Center (LUMC)~\cite{Oreilly_etal}. The gradient coils were tested and used to acquire three-dimensional images of different types of objects. Figure~\ref{fig_3d_scan_results} provides an example of such an image, in which coronal, sagital, and transverse slices through a melon are depicted. Minimal distortion can be observed, which is most likely due to $B_0$ inhomogeneities and not due to any nonlinearities in the gradient fields.

\section{Discussion and Conclusion}
We have applied the target field method to design gradient coils for an MR scanner with a transverse magnetic background field. It is then relatively straightforward to turn these current paths into a constructed gradient coil using simple three-dimensional printing techniques and wire winding. Field measurements confirmed that the proposed design procedure indeed leads to gradient coils that produce the required gradient fields. 

\begin{table*}[tbp]
\caption{Design parameters of the gradient coils. The power is computed for a gradient strength of 10~mT/m.}
\label{tab_tunparam}
\centering
\begin{tabular}{l c c c c c c c c c c}
\toprule
Direction & $n$ & $d$ [mm] & $l$ [m] &  \# turns/quadrant & $a$ [mm]   &$\eta_\mathrm{sim}$ [mT/m/A]  & $\eta_\mathrm{meas}$ [mT/m/A]  & $R$ [$\Omega$]  & $L$ [$\mu$H] &  $P$ [W] \\ \midrule
$z$ & 16 & 140 & 38 & 15 & 135 & 0.52 & 0.59 &0.37 &  180& 329\\
$x$ & 30 & 155 & 42 & 12 & 139 & 0.81 & 0.95 & 0.41 & 227 & 165\\
$y$ & 30 & 155 & 43 & 12 & 137 & 0.82 & 1.02 & 0.40 & 224& 162\\\bottomrule
\end{tabular}
\end{table*}

Transverse $x$- and $y$-gradient coils are generally more efficient, and therefore easier to design with respect to field requirements than the longitudinal $z$-gradient coils. The coil efficiency of the transverse coils is typically less sensitive to the order of the target field profile function and the magnitude of the order is basically limited by the spacing allowed between the wires. The region of uniform linearity is also much larger for transverse gradient coils than for $z$-gradient coils as indicated by the uniformity error that we introduced and the linear spherical volume. The coil efficiency of a $z$-gradient coil, in contrast, strongly depends on the order of the profile function and decreases as the order increases. This indicates that a relatively small order should be chosen to realize an effective $z$-gradient coil, but selecting a small order leads to a very long gradient coil which may be longer than the magnet itself. Careful tuning is therefore necessary to obtain a realizable $z$-gradient coil with a sufficiently large linear region and coil efficiency. Given the cylindrical geometry of our Halbach configuration, it can also be expected that the realization of a $z$-gradient coil in a scanner with a transverse background field is more difficult than realizing transverse gradient coils, since the magnitude of a linearly varying transverse field along the bore of the magnet naturally increases as we move in a radial direction towards the coil. 

Possible extensions of this work include incorporating gradient power minimization as it relates to the Halbach configuration, since this would simplify power supply requirements, which is of importance in a low-resource setting. In addition, using conductive sheets for the construction of a gradient coil may be a feasible large-scale production method.

To summarize, with the proposed design methodology it is possible to design effective $x$-, $y$-, and $z$-gradient coils in case of transverse background fields as encountered in a Halbach permanent magnet scanner. The method is very efficient and allows for parametric coil design, thereby providing insight into the tradeoffs of gradient coil construction.

\appendices

\section{Details of the Modified Target Field Method}
\label{app_tfderiv}
We denote the domain inside the cylinder Region~I, while the domain outside the cylinder is called Region~II. Field quantities having their support in these domains carry a corresponding superscript.

The magnetic field in both domains is governed by the field equations $\nabla \cdot \mathbf{B}=0$ and $\nabla \times \mathbf{B}=\mathbf{0}$. The latter equation is satisfied if we write $\mathbf{B}=-\nabla \Phi$, where $\Phi$ is the scalar magnetic potential. Substitution in the first field equation gives $\nabla^2 \Phi =0$. In other words, the potential satisfies Laplace's equation inside and outside the cylinder. Writing this equation in cylindrical coordinates, we have
\begin{equation}
\label{eq:laplace}
\frac{\partial^2 \Phi^{\text{I,II}}}{\partial r^2} +
\frac{1}{r} \frac{\partial \Phi^{\text{I,II}}}{\partial r} +
\frac{1}{r^2} \frac{\partial^2 \Phi^{\text{I,II}}}{\partial \phi^2} +
\frac{\partial^2 \Phi^{\text{I,II}}}{\partial z^2} =0.
\end{equation}
Furthermore, at the current-carrying surface $r=a$ we have the boundary conditions
\begin{equation}
\label{eq:bcs1}
\lim_{r \uparrow a} \frac{\partial \Phi^{\text{I}}}{\partial r} = \lim_{r \downarrow a} \frac{\partial \Phi^{\text{II}}}{\partial r},
\end{equation}
\begin{align}
\label{eq:bcs2}
\lim_{r \uparrow a} \frac{1}{r} \frac{\partial \Phi^{\text{I}}}{\partial \phi} -
\lim_{r \downarrow a} \frac{1}{r} \frac{\partial \Phi^{\text{II}}}{\partial \phi} = \mu_0 J_z,
\intertext{and}
\label{eq:bcs3}
\lim_{r \downarrow a} \frac{\partial \Phi^{\text{II}}}{\partial z} -
\lim_{r \uparrow a} \frac{\partial \Phi^{\text{I}}}{\partial z} = \mu_0 J_\phi,
\end{align}
and, finally, the surface current must satisfy the continuity equation 
\begin{equation}
\label{eq:div_free}
\frac{\partial J_z}{\partial z} + \frac{1}{a} \frac{\partial J_\phi}{\partial \phi} =0.
\end{equation}

Applying the Fourier transform (\ref{eq:FT_fw}) to Laplace's equation, the boundary conditions, and the continuity equation, we obtain the spectral domain equations
\begin{equation}
\label{eq:Laplace_ft}
r^2 \frac{\partial^2 \tilde{\Phi}^{[m]}}{\partial r^2} +
r \frac{\partial \tilde{\Phi}^{[m]}}{\partial r} -
(m^2 + k^2 r^2) \tilde{\Phi}^{[m]} =0,
\end{equation}
\begin{equation}
\label{eq:bcs1_ft}
\lim_{r \uparrow a} \frac{\partial \tilde{\Phi}^{\text{I};[m]}}{\partial r} = \lim_{r \downarrow a} \frac{\partial \tilde{\Phi}^{\text{II};[m]}}{\partial r},
\end{equation}
\begin{align}
\label{eq:bcs2_ft}
\text{j}m
\left(
\lim_{r \uparrow a} \frac{1}{r} \tilde{\Phi}^{\text{I};[m]} -
\lim_{r \downarrow a} \frac{1}{r} \tilde{\Phi}^{\text{II};[m]}
\right) = \mu_0 \tilde{J}_z^{[m]},
\intertext{and}
\label{eq:bcs3_ft}
\text{j}k
\left(
\lim_{r \downarrow a} \tilde{\Phi}^{\text{II};[m]} -
\lim_{r \uparrow a} \tilde{\Phi}^{\text{I};[m]}
\right)= \mu_0 \tilde{J}_\phi^{[m]}
\end{align}
and
\begin{equation}
\label{eq:div_free_ft}
ka \tilde{J}^{[m]}_z + m \tilde{J}_\phi^{[m]}=0.
\end{equation}
As is well known~\cite{Jin}, the solution of (\ref{eq:Laplace_ft}) in Region~I that is bounded at the origin is given by $\tilde{\Phi}^{\text{I};[m]}(r,k)=\alpha_m(k) I_{m}(|k|r)$, where the coefficient $\alpha_m(k)$ is independent of $r$, while the solution in Region~II that remains bounded as $r \rightarrow \infty$ is given by $\tilde{\Phi}^{\text{II};[m]}(r,k)=\beta_m(k) K_{m}(|k|r)$ with $\beta_{m}(k)$ independent of $r$. Substituting these solutions in the boundary conditions, the coefficients are found as
\begin{align}
\label{eq:alpha}
\alpha_m(k) &=
-\text{j} a \mu_0 \frac{|k|}{k} K'_m(|k|a) \tilde{J}_\phi^{[m]}
\intertext{and}
\label{eq:beta}
\beta_m(k) &= -\text{j} a \mu_0 \frac{|k|}{k} I'_m(|k|a)\tilde{J}_\phi^{[m]}.
\end{align}
Having the spectral domain potential at our disposal, the corresponding spectral domain magnetic field can be determined. Of particular interest is the magnetic field inside the cylinder (Region~I), since the target field is prescribed in this region. Explicitly, for the magnetic field in Region~I, we have
\begin{align}
\label{eq:Brsp_I}
\begin{split}
\tilde{B}_r^{\text{I};[m]} &= -
\frac{\partial \tilde{\Phi}^{\text{I};[m]}}{\partial r} \\
&= \text{j} a \mu_0 k I'_m(|k|r) K'_m(|k|a) \tilde{J}_\phi^{[m]},
\end{split}
\end{align}
\begin{align}
\label{eq:Bphisp_I}
\begin{split}
\tilde{B}_\phi^{\text{I};[m]} &= -\frac{\text{j}m}{r} \tilde{\Phi}^{\text{I};[m]} \\
&= -\frac{a\mu_0}{r} m \frac{|k|}{k} I_m(|k|r) K'_m(|k|a) \tilde{J}_\phi^{[m]},
\end{split}
\end{align}
and
\begin{align}
\label{eq:Bzsp_I}
\begin{split}
\tilde{B}_z^{\text{I};[m]} &= - \text{j}k \tilde{\Phi}^{\text{I};[m]} \\
&= -a \mu_0 |k| I_m(|k|r)K_m'(|k|a) \tilde{J}_\phi^{[m]}.
\end{split}
\end{align}
Now the target field is in the $x$-direction and is prescribed on the inner cylinder $r=b$ . Writing this field in terms of its cylindrical components, we have
\begin{equation}
\label{eq:Bx_cyl}
B_x(b,\phi,z) = B_r^{\text{I}}(b,\phi,z) \cos(\phi) - B_\phi^{\text{I}}(b,\phi,z) \sin(\phi)
\end{equation}
and applying the Fourier transform to the above equation gives
\begin{align}
\label{eq:Bx_cyl_ft}
\begin{split}
\tilde{B}_x^{[m]}(b,k) &=
\frac{1}{2}
\left[
\tilde{B}_r^{\text{I};[m-1]}(b,k) + \tilde{B}_r^{\text{I};[m+1]}(b,k)
\right] \\
&- \frac{1}{2\text{j}}
\left[
\tilde{B}_\phi^{\text{I};[m-1]}(b,k) - \tilde{B}_\phi^{\text{I};[m+1]}(b,k)
\right].
\end{split}
\end{align}
Substituting (\ref{eq:Brsp_I}) and (\ref{eq:Bphisp_I}) in the above expression we arrive at (\ref{eq:Bx_F}).

\section{Surface current density for a $z$-gradient coil}
\label{app_graddesign}
We show how we obtain the surface current from the prescribed target field for the design of a $z$-gradient coil. The analysis for an $x$- or $y$-gradient coil runs along similar lines.

For a $z$-gradient coil, the Fourier transform of the target field is given by $\tilde{B}_x^{[m]}(b,k) = 2\pi g_z \tilde{\Gamma}_\text{ln}(k) \delta_{m,0}$, where the delta symbol denotes the Kronecker delta and
\begin{equation}
\label{eq:Gamma_ln_ft}
\tilde{\Gamma}_\text{ln}(k)  =
\int_{z=-\infty}^{\infty} \Gamma_{\text{ln}}(z) e^{-\text{j}kz} \, \text{d}z.
\end{equation}
Note that $\tilde{\Gamma}_{\text{ln}}(k)$ is imaginary and an odd function of $k$.

Substitution of the Fourier transform of the target field in (\ref{eq:Bx_F}) gives
\begin{align}
\label{eq:Bx_F_zg}
\begin{split}
& 2\pi g_z \tilde{\Gamma}_\text{ln}(k) \delta_{m,0} = \\
&\frac{\text{j}}{2}
\left[
\tilde{P}^{[m-1]}(b,k) - \tilde{Q}^{[m-1]}(b,k)
\right]
\tilde{J}^{[m-1]}_\phi(k)\\
&+
\frac{\text{j}}{2}
\left[
\tilde{P}^{[m+1]}(b,k) + \tilde{Q}^{[m+1]}(b,k)
\right]
\tilde{J}^{[m+1]}_\phi(k).
\end{split}
\end{align}

Since the left-hand side of this equation vanishes for $m$ odd, we take a surface current for which all even numbered angular modes of its $\phi$-component vanish, that is, we take $\tilde{J}_\phi^{[m]}(k)=0$ for $m$ even and $k \in \mathbb{R}$. Furthermore, for $m=0$ we obtain
\[
2\pi g_z \tilde{\Gamma}_\text{ln}(k) = \frac{\text{j}}{2}
\left[
\tilde{P}^{[1]}(b,k) + \tilde{Q}^{[1]}(b,k)
\right]
\left[
\tilde{J}_\phi^{[-1]}(k) + \tilde{J}_\phi^{[1]}(k)
\right],
\]
where we have taken the symmetry of $\tilde{P}^{[m]}$ and $\tilde{Q}^{[m]}$ with respect to $m$ into account. For the surface current we now take $\tilde{J}_\phi^{[-1]}(k)=\tilde{J}_\phi^{[1]}(k)$ and we obtain
\begin{equation}
\label{eq:mode1}
\tilde{J}_\phi^{[1]}(k) = - \text{j}\frac{2\pi g_z \tilde{\Gamma}_{\text{ln}}(k)}{\tilde{P}^{[1]}(b,k) + \tilde{Q}^{[1]}(b,k)} =
\tilde{J}_\phi^{[-1]}(k).
\end{equation}
Similarly, for $m$ even and not equal to zero ($m=2n$, $n= \pm 1, \pm 2, ..$) the left-hand side vanishes and if we take a surface current for which all odd numbered angular modes are even with respect to $m$, that is,
\begin{equation}
\label{eq:sym}
\tilde{J}_\phi^{[-2n+1]}(k) = \tilde{J}_\phi^{[2n-1]}(k), \quad n=1,2,...,
\end{equation}
then we satisfy (\ref{eq:Bx_F_zg}) if
\begin{equation}
\label{eq:rec}
\tilde{J}_\phi^{[2n+1]}(k) = -
\frac{ \tilde{P}^{[2n-1]}(b,k) - \tilde{Q}^{[2n-1]}(b,k) }{ \tilde{P}^{[2n+1]}(b,k) + \tilde{Q}^{[2n+1]}(b,k) }
\tilde{J}_\phi^{[2n-1]}(k),
\end{equation}
for $n=1,2,...\,$.
In other words, all odd-numbered higher-order modes can be determined recursively starting from $\tilde{J}_\phi^{[1]}(k)$ as given by (\ref{eq:mode1}).

To obtain the $\phi$-component of the surface current in the spatial domain, we substitute the modes in the Fourier inversion formula and include apodization to obtain
\begin{align}
\label{eq:Jphi_spatial}
\begin{split}
J^z_\phi(\phi,z) &=
\frac{1}{4\pi^2}
\int_{k=-\infty}^{\infty}
\sum_{m=-\infty}^{\infty}
\tilde{J}_\phi^{[m]}(k) \tilde{T}(k) e^{\text{j}m\phi} e^{\text{j} k z} \, \text{d} k \\
&=
\frac{1}{2\pi^2}
\sum_{\underset{\text{$m$ odd}}{m=1}}^{\infty}
\cos(m\phi)
\int_{k=-\infty}^{\infty}
\tilde{J}_\phi^{[m]}(k) \tilde{T}(k)
e^{\text{j} k z} \, \text{d} k.
\end{split}
\end{align}
The current consists of an infinite summation of odd-numbered angular modes. Each term in this series represents the $\phi$-component of a surface current that produces its own magnetic field. The total magnetic field consists of a superposition of these individual fields due to the linearity of the field equations. Since we want to realize a $z$-gradient coil in practice, we have to truncate the series and to keep the construction of the coil as simple as possible, we keep the first current term in the series only. Our final expression for the $\phi$-component of the surface current becomes
\begin{align}
\label{eq:Jphi_spatial_final}
\begin{split}
J^z_\phi(\phi,z) &=
\frac{1}{2\pi^2}
\cos(\phi)
\int_{k=-\infty}^{\infty}
\tilde{J}_\phi^{[1]}(k) \tilde{T}(k)
e^{\text{j} k z} \, \text{d} k \\
&=
-\text{j} \frac{g_z}{\pi} \cos(\phi)
\int_{k=-\infty}^{\infty}
\frac{\tilde{\Gamma}_{\text{ln}}(k)\tilde{T}(k)}{ \tilde{P}^{[1]}(b,k)+\tilde{Q}^{[1]}(b,k) }
e^{\text{j} k z} \, \text{d} k.
\end{split}
\end{align}
The $z$-component of the surface current that corresponds to (\ref{eq:Jphi_spatial_final}) follows from the continuity equation for the surface current.

\section*{Acknowledgment}
The authors would like to thank Wouter Teeuwisse for fruitful discussions and his help in the construction and assembly of the gradient coils, Danny de Gans for his work on the gradient amplifier, and Martin van Gijzen for his support in the low field MRI scanner project. This research was made possible in part through the NWO-WOTRO grant  nr. W 07.303.101:`A sustainable MRI system to diagnose hydrocephalus in Uganda', Horizon 2020 ERC Advanced NOMA-MRI 670629: and the Simon Stevin Prijs from the NWO.

\end{document}